# An Artificial Intelligence Framework for Joint Structural-Temporal Load Forecasting in Cloud Native Platforms

Qingyuan Zhang
Boston University
Boston, USA

*Abstract-This study targets cloud native environments where microservice invocation relations are complex, load fluctuations are multi-scale and superimposed, and cross-service impacts are significant. We propose a structured temporal joint load prediction framework oriented to microservice topology. The method represents the system as a coupled entity of a time-evolving service invocation graph and multivariate load sequences. It constructs neighborhood-aggregated and global summarized views based on service level observations. This forms layered load representations across instance, service, and cluster levels. A unified sequence encoder models multi-scale historical context. To strengthen the expression of invocation dependencies, the framework introduces a lightweight structural prior into attention computation. This enables more effective capture of load propagation and accumulation along invocation chains, while maintaining consistent modeling of local bursts and overall trends. The training objective adopts a multi-objective regression strategy that jointly optimizes service level and cluster level predictions to improve cross-granularity stability. We further conduct single-factor sensitivity analyses on key structural and training hyperparameters. We systematically examine the effects of time window length, encoding depth, and regularization strength. The results support the necessity of multi-granularity fusion and structural injection and clarify their effective configuration ranges. Overall, the framework provides a reusable modeling paradigm and implementation path for capacity assessment, resource orchestration, and runtime situational understanding in cloud environments.*

## I. Introduction

Cloud native architectures are reshaping computing and service delivery. Under the container-based scheduling paradigm led by Kubernetes, applications shift from traditional monolithic systems to complex structures composed of many loosely coupled microservices. This architecture enables elastic scaling, continuous delivery, and rapid iteration. It also increases the difficulty of resource scheduling and service governance. In real production environments, microservice instances are highly dynamic. Business traffic fluctuates across time and space. Resource demands change rapidly with scenarios, user behavior, and dependency chains. Understanding and forecasting such complex load evolution is essential for reducing overprovisioning, avoiding bottlenecks, and improving service stability. Load prediction affects not only the resource needs of an individual service but also global scheduling decisions, cross-layer scaling strategies, and overall service quality[1].

Existing load prediction methods mainly rely on statistical modeling or deep learning for time series. However, they cannot often model the dependency structure of cloud native microservices. In microservice architectures, service invocation forms complex topologies. Load changes frequently arise from cross-service propagation. For example, heavy traffic on upstream services propagates downstream. Database load depends on query patterns. Message queues may be overwhelmed by sudden bursts. Traditional methods treat loads as isolated time series and ignore topology dependencies and semantic propagation. This leads to weak prediction performance and low sensitivity to anomalies. In addition, cloud native loads exhibit multi-granularity characteristics. At the granularity level, there are instance level, microservice level, node level, and cluster level. At the temporal level, there are second-level bursts and hour-level trends. At the semantic level, there are dynamic couplings among different request types. These multidimensional and multiscale characteristics make simple models insufficient for accurate prediction[2].

In this context, introducing topology awareness and representation learning is of great importance. Microservice topology is not a static graph. It evolves with deployment, scaling, migration, and container scheduling. Load prediction must capture local dependencies and model cross-path propagation. Explicit modeling of such a structure helps reveal influence directions and magnitudes among services. It improves sensitivity to dynamic changes and enhances predictive foresight. Topology information also supports anomaly diagnosis, root-cause tracing, and capacity planning. Therefore, load prediction in cloud native systems should not be treated as a pure time series problem. It should be viewed as a task involving deep coupling between topology structure and temporal dynamics[3].

Meanwhile, Transformer models have demonstrated strong global dependency modeling capability for sequential representations. This provides a new perspective for multi-granularity load prediction in cloud native systems. Transformers capture long-range dependencies without the constraints of recurrent connections. This makes them suitable for long temporal links caused by complex service chains. However, conventional Transformers lack inherent awareness of topology and cannot directly fuse multi-granularity information. Mechanism-level innovations are needed. In cloud native scenarios, employing Transformers to model multi-granularity load sequences and integrating explicit topology modeling allows the model to understand cross-service relationships. This can significantly improve accuracy and robustness[4].

From a system perspective, high-quality load prediction affects resource scheduling strategies, cost, reliability, and user experience. By analyzing service topology and multi-granularity load patterns, it is possible to implement fine-grained scheduling and reduce resource waste. This enhances system availability and alleviates delays and failures under burst pressure. Multi-granularity prediction supports elastic scaling that adapts smoothly to peak and off-peak conditions. This is valuable for handling complex workloads, improving service elasticity, and reducing operational cost. Therefore, constructing a topology-aware deep modeling scheme for load prediction has essential theoretical significance and broad application prospects. It forms a foundation for intelligent operations, automated scheduling, and autonomous systems in cloud native environments.

## II. Related work

In recent years, load prediction in cloud and edge computing environments has received extensive attention. The research focus has shifted from traditional statistical models to deep learning models in order to better capture complex and dynamic resource needs. Zheng et al. proposed the MC TIDE model. It applies multi-channel dense encoding for multivariate load data and extracts deep features. This allows the model to capture interactions among different resource indicators. Their results show that fusing multi-channel features has important value for data center energy optimization[5]. Joha et al. combined real-time short-term load prediction with anomaly detection in industrial IoT settings. They developed a self-adaptive system with multiple cooperating modules. The mechanism effectively improves prediction and diagnostic performance[6]. These studies indicate that load prediction has gradually evolved from a pure time series task to a system-level strategy for resource management and risk mitigation.

With respect to algorithm selection, some research compares the advantages of different deep learning structures. Zhang et al. proposed a CNN BiLSTM-based method for cloud load prediction. The convolution structure extracts local temporal patterns. The bidirectional LSTM then models complex sequence dependencies. The method significantly improves prediction accuracy[7]. Guruge et al. combined LSTM with Facebook Prophet for Kubernetes auto scaling. The method integrates time trend modeling with deep sequence learning. It makes scaling strategies more stable and forward-looking[8]. These works demonstrate the feasibility of hybrid structures in load prediction. However, they still focus mainly on one-dimensional time sequence features and lack deep modeling of service topology.

Some scholars investigate load prediction from the perspective of distributed environments. Sharma and Kaur proposed a federated time series prediction framework for IoT based on fog computing. It enables localized data computation and collaborative modeling. The framework reduces communication cost and mitigates privacy risks in distributed tasks[9]. Ding et al. proposed a nonlinear time series-based method to model server status. The core idea is to capture latent nonlinear patterns of system operation. This improves the ability to anticipate load trends[10]. These studies show that load prediction involves not only model accuracy but also data distribution, computation location, and cooperation mechanisms. Traditional single-node strategies are becoming less suitable.

In addition, some studies address finer-grained resource management in cloud native settings. For example, Lin proposed a time series-based method to predict cold start time in serverless platforms. By modeling function trigger behavior, the approach reduces the performance impact of cold start latency[11]. These works revisit the practical value of load prediction from the perspective of system operation policies. They emphasize the close connection between dynamic resource scheduling and forecasting. However, they still focus mainly on time series modeling. They pay limited attention to complex microservice topology, cross-service dependency propagation, and multi-granularity information fusion. A system-level and topology-aware predictive framework is still lacking, which provides a clear direction for further research.

## III. Method

We represent a cloud-native microservice system as a combination of a time-varying call topology and a multi-granularity observation sequence. Let V be the set of microservices within a time window, and E be the call relationships; then the topology can be represented as a directed graph $G = (V, E)$. An adjacency matrix $A \in R^{V \times V}$ describes the call intensity or the existence of calls between services, where $A_{ij} = 1$ indicates that a call exists from service i to service j, or the call intensity is represented by normalized weights. The load observation vector for each service at time t is denoted as $x_{ij}$, and d can include basic metrics such as CPU, memory, QPS, and latency. To unify different metric scales, we perform simple standardization on the input:

$$\widetilde{x}_t = \frac{x_t - \mu}{\sigma} \qquad (1)$$

Here, $\mu, \sigma$ represents a sliding window or global statistic, enabling the model to more stably learn the relative variation patterns of different microservice loads. Its model architecture is shown in Figure 1.

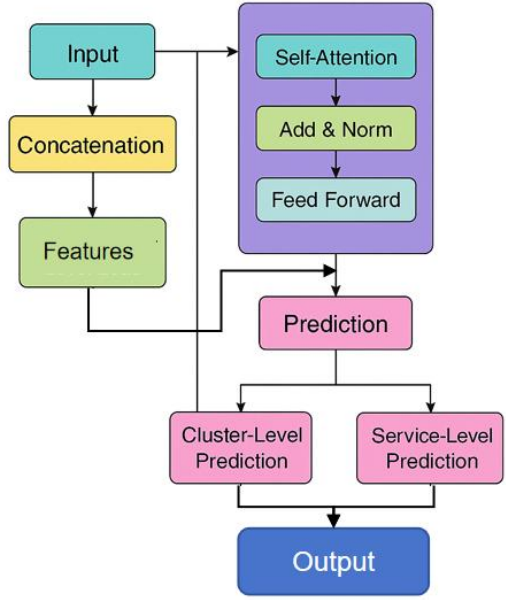

Figure 1. Overall Algorithm Architecture Diagram

To achieve multi-granularity modeling, we construct sequence views at the instance, service, and cluster levels simultaneously, and use lightweight aggregation to ensure scalability. Let the service-level input be $\tilde{x}_t$, then the cluster-level load can be obtained by summation or averaging:

$$x_t = \frac{1}{|V|}\sum_{i \in V} \tilde{x}_t^{(i)} \quad (2)$$

Simultaneously, to inject local topological context, we construct a first-order neighborhood aggregation feature for each service:

$$h_t^{(i)} = \sum_{j \in V} A_{ij} \tilde{x}_t^{(i)} \quad (3)$$

Therefore, the multi-granularity fusion input of each service at time t can be written as a simple concatenation:

$$z_t^{(i)} = Concat(\tilde{x}_t^{(i)}, h_t^{(i)}, g_t) \quad (4)$$

This design uses minimalist graph neighborhood information to characterize the cross-service propagation trend of load, while providing system situation background with global granularity, enabling the same Transformer to perceive the three layers of load semantics: "local-neighborhood-global".

For temporal encoding, we employ a Transformer-based sequence modeling structure and incorporate topology-aware attention bias to reflect invocation dependencies. For any service i, we map $z_t^{(i)}$ over the most recent T time steps to query, key, and value:

$$Q = ZW_Q, \ K = ZW_K, \ V = ZW_V \quad (5)$$

Where Z is the time-series stacking matrix of the service. To make the attention more biased towards service patterns that have strong dependencies, we add a simple topological bias to the attention scoring:

$$Attn(Q,K,V) = Softmax(\frac{QK^T}{\sqrt{d}} + B)V \quad (6)$$

Here, $B$ can be obtained from the topology or its low-order transformations. For example, for the sequence of services i, it can be generated using constant or lightweight mappings related to their adjacency strength, allowing the model to retain awareness of "call chain-driven load propagation" while learning long-range temporal dependencies. This form of attention keeps the computational structure simple, does not introduce complex graph operators, and is easy to deploy in large-scale microservice clusters.

The prediction layer employs multi-task outputs to correspond to multi-granularity objectives. Let $y_{t+1:t+H}$ represent the service-level prediction of H steps ahead, and $g_{t+1:t+H}$ represent the cluster-level prediction. We use a simple linear header to connect to the last layer of the Transformer representation and employ weighted mean squared error as the training objective:

$$L = \alpha \|\widehat{y} - y\|^2 + \beta \|\widehat{g} - g\|^2 \quad (7)$$

Here, $\alpha, \beta$ is used to balance the importance of local and global predictions. This objective design allows the model to learn the fine fluctuations of individual services while maintaining a stable grasp of the overall system trend, thus providing a unified and interpretable multi-granularity load prediction basis for subsequent elastic scaling, capacity planning, and congestion prevention.

IV. EXPERIMENTAL RESULTS

A. Dataset

This study uses the open microservice tracing data from the Alibaba Cluster Trace Program as the research foundation. We specifically select the dataset cluster trace microservices v2021. The dataset originates from a real production environment and spans a 12-hour window. It contains observations of more than 20,000 microservices and provides dependency clues and key performance and load statistics such as invocation dependencies, response time, and request rate. It offers both structural and temporal evidence for modeling load evolution and cross-service propagation in cloud native microservices. The dataset is released in a public repository, ensuring reproducibility and comparability. It is well-suited for topology-aware load prediction research in microservice environments.

At the data organization level, we treat each microservice as a node to construct service-level sequences. Based on invocation dependency information, we extract directed relations among services and form a call topology representation corresponding to each time window. For the load dimension, we construct multivariate time series using time-varying metrics such as invocation rate and response time. These features, combined with the topology, produce a joint structure time input. To support multi-granularity forecasting, we further aggregate service level sequences to obtain a system-level load view. This allows the model to learn both local service fluctuations and global dynamics, aligning with the multi-granularity prediction objective.

The dataset has clear advantages. It reflects a realistic scale, it provides explicit topology information, and it closely couples invocation structure with runtime indicators. It supports modeling of dependency propagation, hotspot concentration, and link-level fluctuations in microservice workloads. Compared with traditional datasets that only contain host-level or cluster-level resource utilization, this tracing dataset captures the essential nature of service invocation in cloud native architectures. It is more suitable for evaluating topology-aware sequence modeling. Based on this data foundation, the proposed framework can jointly describe service level and cluster-level load in a unified data context. It provides structurally grounded input for intelligent scheduling and elastic management in cloud native systems.

### B. Experimental setup

The experiments are conducted on a single machine with a single GPU. The hardware configuration includes an NVIDIA A100 80 GB GPU, an Intel Xeon Gold 6338 CPU with 32 cores, 256 GB memory, and a 2 TB NVMe SSD. The software environment consists of Ubuntu 20.04, Python 3.10, PyTorch 2.1.0, CUDA 12.1, and cuDNN 8.9. The model uses a Transformer-based topology-aware multi-granularity prediction structure. The sequence length is 60, and the prediction horizon is 12. The input feature dimension is aligned at the service level. The Transformer encoder has 4 layers, 8 attention heads, a hidden dimension of 256, a feed-forward dimension of 1024, and dropout of 0.1. AdamW is used as the optimizer with an initial learning rate of 1e-4, weight decay 1e-2, batch size 64, and 50 training epochs. Gradient clipping is applied with a threshold of 1.0. Linear warmup is applied for the first 5 percent of training steps, followed by cosine decay. The loss function is a weighted mean squared error combining service level and system level predictions, with weights 0.7 and 0.3. To ensure reproducibility, the random seed is set to 42. The dataset is split chronologically into training, validation, and testing sets with a ratio of 8 to 1 to 1. All inputs are standardized using statistics computed on the training set.

### C. Experimental Results

This paper first conducts a comparative experiment, and the experimental results are shown in Table 1. Here, R-Score denotes the coefficient of determination between the predicted and ground-truth loads, where a higher value indicates better regression fit.

Table 1. Comparative experimental results

| Model | MSE | MAE | MAPE | R-Score |
|---|---|---|---|---|
| Etsa-lp[12] | 0.0282 | 0.1174 | 6.42% | 0.927 |
| Temposcale[13] | 0.0229 | 0.1031 | 5.87% | 0.941 |
| AmazonAICloud[14] | 0.0196 | 0.0978 | 5.21% | 0.953 |
| Ours | 0.0143 | 0.0856 | 4.38% | 0.965 |

The results show that the proposed topology-aware multi-granularity load prediction framework outperforms all comparison models on every evaluation metric. This indicates that traditional time series-based methods are unable to fully capture cross-service dependencies and propagation patterns in cloud native microservice systems. When the microservice topology is complex and the request path varies significantly, the explicit modeling of service relationships reduces prediction bias and makes trend estimation more stable and forward-looking.

For the MSE and MAE metrics, the improvements reflect a stronger capability in fitting load variation. The reductions compared with baselines are significant. This shows that incorporating topology features not only improves short-term fluctuation fitting but also enhances sensitivity to local changes. Traditional models often rely on short sliding windows of historical values. The use of multi-granularity fusion in the proposed framework alleviates misjudgment under burst loads and improves the modeling of load propagation along service chains.

The MAPE metric reflects overall prediction stability. The improvement in this metric indicates stronger adaptability to different scenarios and service types. It suggests better generalization capability. In complex cluster runtime states, load variation may be influenced by routing policies, invocation chain propagation, and evolving service topology. The topology-aware mechanism strengthens structural understanding when modeling correlations and helps reduce errors arising from changes in workload characteristics.

The R-Score metric further shows stronger capability in capturing global trends. The proposed method enhances local accuracy while maintaining the smoothness and reliability of system-level prediction. This has practical value for elastic scaling decisions in cloud native environments. It enables earlier detection of potential bottlenecks and supports proactive actions. Overall, the comparison results demonstrate the value of combining topology and multi-granularity modeling in microservice based cloud native systems.

The time window length T determines the range of historical contexts that the model can "see," and is one of the most critical time-series hyperparameters in cloud-native microservice load prediction. For the Transformer framework, which is geared towards topology awareness and multi-granularity modeling, T affects both the ability to capture short-term bursts and local fluctuations, and the depth of understanding of cross-service call chain load propagation and medium- to long-term trends. If T is too short, the model may not be able to fully utilize the historical collaborative changes of topology-related services; if T is too long, it may introduce redundant noise and weaken the focus on the current state. Therefore, it is necessary to conduct a single-factor sensitivity analysis on the time window length while keeping the method in this paper unchanged, to evaluate the stability and adaptation range of the model under different historical context settings, and provide a more reproducible hyperparameter basis for online deployment and elastic scheduling in cloud-native scenarios. The experimental results are shown in Figure 2.

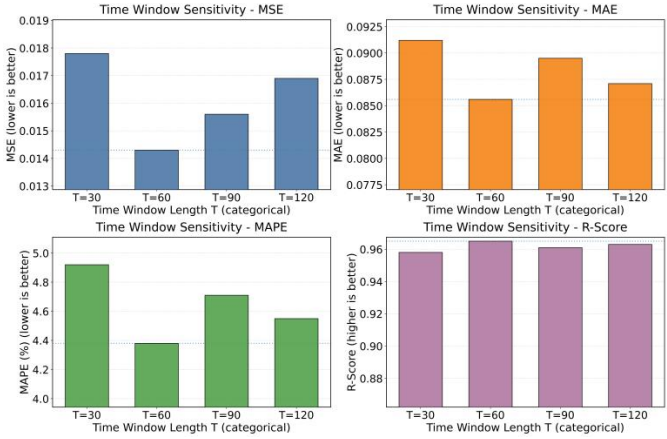

Figure 2. Sensitivity experiment with time window length T

The single-factor sensitivity curve for the time window length T shows that model performance depends strongly on the range of historical context, but longer windows are not always better. In cloud native microservice environments, load is influenced not only by the temporal pattern of each service but also by propagation and accumulation along the invocation topology. Thus, an appropriate value of T reflects a balance between capturing sufficient propagation history and avoiding excessive stale noise. The curve indicates that the proposed framework has an optimal interval for using temporal context, which aligns well with the topology-aware multi-granularity prediction mechanism.

For the error metrics, MSE, MAE, and MAPE reach better performance when T is around 60. When T is set to 30, errors are clearly larger. This suggests that a short window fails to include key propagation segments in microservice chains. In particular, under common cloud native patterns such as upstream triggering and downstream amplification or asynchronous queue buffering, the model lacks enough historical evidence to distinguish between transient spikes and true trends. When T increases to 90 or 120, the errors rise again. This indicates that an overly long history weakens the relevance of current conditions and causes the Transformer's attention to be distracted by weak correlations from early timestamps.

The R-Score trend is consistent with the error metrics. The model achieves a higher overall fit when T equals 60. This means that the model not only better predicts local fluctuations but also more reliably captures joint trends at both service and system levels. This simultaneous improvement in both local error reduction and global trend fitting reflects the benefit of multi-granularity fusion under a proper window. The model preserves short-term behavioral features of individual services and stabilizes trend estimation through topology neighborhood information and global aggregation signals. This reduces the influence of transient disturbances that frequently occur in cloud native environments.

From an engineering and deployment perspective, this sensitivity result guides online load prediction and elastic scheduling. Setting T within a moderate range yields a practical trade-off among cost, timeliness, and accuracy. For the topology-aware Transformer framework, T equal to 60 acts as an effective memory length that matches the propagation scale of service invocations. It improves the interpretability and stability of link-level load correlation. It is also suitable as a default or recommended window setting for real clusters. In addition, this choice reduces the need for frequent retuning when traffic patterns shift, which simplifies long-running production maintenance. A moderate T helps control feature staleness while avoiding excessive truncation of multi-hop dependency signals, making the model more robust to transient spikes. This further supports predictable latency and resource overhead during inference, which is critical for stable autoscaling decisions in high-concurrency microservice environments.

The number of Transformer encoder layers, L, directly determines the model's representation depth and its ability to model cross-temporal dependencies, making it a key structural hyperparameter affecting the effectiveness of topology-aware load prediction. For cloud-native microservice scenarios, an appropriate number of layers needs to balance the complexity of information transfer in the service call topology with the multi-scale fluctuation characteristics of the load sequence. Therefore, it is necessary to conduct a single-factor sensitivity analysis on L while maintaining the framework and data processing flow of this paper, to evaluate the model's stability and deployability at different structural depths. The experimental results are shown in Figure 3.

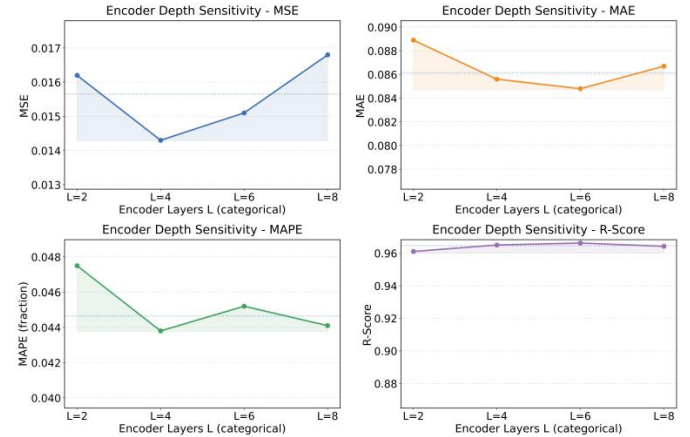

Figure 3. The impact of the number of Transformer encoder layers on experimental results

The sensitivity curve for the encoder layer number L shows that structural depth has a clear impact on topology-aware multi-granularity load prediction, and a moderate depth appears to be more robust. In cloud native microservice systems, load variation is not only temporal fluctuation within a single service. It also includes correlated changes propagated along the invocation topology. The model, therefore, needs enough layers to aggregate useful information across time and across services. If the depth is too shallow, the representation of structure and temporal coupling becomes limited. The model then struggles to capture synchronized changes between link-level interactions and global trends.

The error trends indicate that L=4 and L=6 achieve better and more balanced performance. Under the current fusion input of service level, neighborhood level, and cluster level signals, a moderate depth can effectively absorb topology priors and filter historical context. In contrast, when L increases to 8, MSE and MAE rise again. This suggests that an overly deep structure may amplify noise or cause excessive smoothing. It may also weaken the focus of attention allocation across multi-granularity information. As a result, the model becomes less sensitive to short-term bursts and local jitter in microservice loads.

The R-Score remains relatively high with small fluctuations across different L values. This implies that the proposed framework has a degree of structural robustness in trend modeling. However, fine-grained errors still change with depth. From an engineering perspective in cloud native environments, a moderate number of layers offers a more practical balance among prediction accuracy, training stability, and online inference cost. It also matches the real requirement of topology-aware load prediction. The model should have sufficient expressive power without unnecessary complexity. Therefore, a moderate depth is a safer structural choice for subsequent multi-granularity elastic scheduling and capacity planning.

The dropout ratio is a key regularization hyperparameter affecting the generalization ability and training stability of Transformers, especially in scenarios with high noise, strong fluctuations, and significant topological dependencies, such as cloud-native microservice loads. For a topology-aware multi-granularity prediction framework, a suitable dropout needs to strike a balance between preserving cross-service dependency representation capabilities and suppressing overfitting. Therefore, while maintaining the model structure and data processing flow in this paper, a single-factor sensitivity analysis of the dropout ratio helps to clarify the robustness boundaries and engineering usability range of the framework under different regularization strengths. The experimental results are shown in Figure 4.

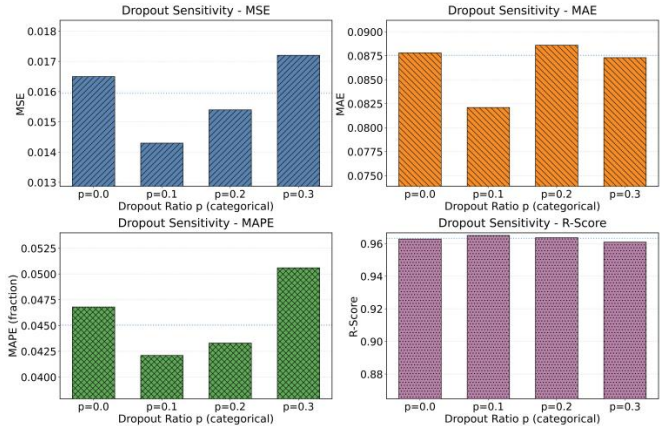

Figure 4. The effect of the dropout ratio on experimental results

The figure shows that the Dropout rate has a significant effect on the proposed topology-aware multi-granularity Transformer load prediction framework, and the performance changes nonlinearly with regularization strength. In cloud native microservice systems, loads are driven by both temporal dependence and invocation topology propagation. The model needs enough representation capacity to capture cross-service coupling. It also needs appropriate regularization to suppress noise memorization in dynamic environments. In this sense, Dropout serves as a stabilizer for the joint structure and temporal representation.

For the error metrics, MSE, MAE, and MAPE achieve better values when p is around 0.1. When p equals 0.0, the overall errors are higher. This suggests that removing Dropout makes the model more likely to be dominated by local bursts, short-term jitter, or link-level noise. This risk becomes stronger after multi-granularity fusion. Excessive fitting can amplify the influence of non-critical historical segments. A moderate Dropout rate helps attention focus on stable topology collaboration patterns and transferable temporal structures. It also improves consistency between service level and cluster level predictions.

When Dropout increases to p=0.2 or even p=0.3, the errors rise noticeably. This indicates that overly strong random deactivation weakens the effective use of key dependency paths and historical context. In microservice topologies with load propagation chains, an overly high Dropout rate may disrupt cross-layer semantic continuity. Neighborhood-level and global-level information can be overly diluted during encoding. This then harms the estimation of true load evolution trends and reduces the ability to distinguish burst peaks from normal fluctuations.

The R-Score trend corroborates the error metrics. The overall fit is better at p=0.1, while the fitting ability declines at the two extremes. In summary, the results indicate a clear effective range of regularization strength for the proposed method in cloud native microservice scenarios. A moderate Dropout rate enables a more robust representational balance between complex topology and multiscale load dynamics. This also provides a reliable hyperparameter reference for elastic scaling in online deployments. It improves robustness to environmental changes while maintaining prediction accuracy.

## V. CONCLUSION

This paper addresses key challenges in cloud native microservice environments, including complex load patterns, strong invocation chain dependencies, multiple granularity layers, and rapid dynamics. We propose a Transformer-based topology-aware multi-granularity load prediction framework. The framework is grounded in real system operation logic. It brings the coupling between microservice invocation topology and time series loads into a unified modeling view. By jointly fusing service level, neighborhood level, and cluster level information, it strengthens the ability to model cross-service load propagation, local burst disturbances, and global trend evolution. Compared with traditional paradigms that focus on a single sequence or a single granularity, the proposed method better matches the structural nature of cloud native systems. It provides a clear and reusable technical route and methodological support for shifting load prediction from time-only modeling to joint structure and time modeling.

From the perspective of engineering value and practical impact, this work offers a structurally grounded prediction basis for intelligent operations and autonomous scheduling in cloud native systems. Accurate and stable multi-granularity consistent load prediction can improve the reliability of elastic scaling decisions. It can reduce over-scaling and resource waste. It can also identify potential risks earlier under complex conditions such as traffic fluctuations, hotspot migration, and invocation rerouting. This reduces the probability of service jitter, latency escalation, and cascading failures. Moreover, topology-aware prediction can provide invocation chain-oriented priors for capacity planning, SLA assurance, anomaly warning, and root cause analysis. This supports a shift in cloud native governance from experience-driven practice to data-driven and structure-driven decision-making. It brings more direct benefits for stability, cost control, and performance optimization in large-scale clusters.

This study also shows that single-factor sensitivity analysis of key hyperparameters and environment-related settings is practically useful. Fine-grained analysis of time window length, encoder layer number, and Dropout rate reveals more suitable structural depth and historical context range for topology-aware Transformer frameworks. It also identifies a regularization strength interval that better fits dynamic cloud native scenarios. These findings enhance interpretability and deployability. They provide clear engineering guidance for future work. Improving prediction performance in complex systems should not rely on blindly increasing model size or history length. It should instead focus on structured hyperparameter design guided by topology propagation scales, load fluctuation frequencies, and multi-granularity coordination objectives.

Looking ahead, cloud native microservices will continue to evolve, and load prediction will move from a standalone capability to closed-loop intelligence tightly coupled with scheduling strategies. Future work may explore more refined modeling of dynamic topologies. This can make the model more sensitive to real changes, such as service decomposition, canary releases, instance migration, and fault isolation. The prediction targets can also be extended to finer link-level or request-type level loads. This would improve the modeling of business semantics-driven variations. At the system level, it is valuable to jointly design prediction modules with adaptive scaling, energy optimization, and multi-objective SLA constraints. This can build deployable integrated prediction and decision frameworks and further advance autonomous cloud native platforms. Overall, this work provides an extensible and engineering-ready foundation for topology-aware and multi-granularity load prediction. It is expected to have a sustained influence on intelligent scheduling, cloud energy saving, stable operation of critical services, and the development of next-generation autonomous cloud native systems.